\documentclass[aps,prd,twocolumn,nofootinbib,superscriptaddress]{revtex4-2}
\usepackage{ae}
\usepackage{bm} 
\usepackage{color}
\usepackage{amsmath}
\usepackage{amssymb}
\usepackage{amsfonts}
\usepackage{graphicx}
\usepackage{setspace}
\usepackage{stackrel}
\usepackage{ulem}
\usepackage{braket}
\usepackage{xfrac}
\usepackage{siunitx}
\usepackage{upgreek}
\usepackage{hyperref}
\usepackage{scrextend}
\usepackage{yfonts}

\graphicspath{{./plots/}}

\newcommand{\Eqref}[1]{Eq.~\eqref{#1}}

\allowdisplaybreaks

\begin{document}

\title{Limitations of the paraxial beam model in the study of quantum vacuum signals}
\author{Felix Karbstein}\email{felix.karbstein@uni-jena.de}
\affiliation{Helmholtz-Institut Jena, Fr\"obelstieg 3, 07743 Jena, Germany}
\affiliation{GSI Helmholtzzentrum f\"ur Schwerionenforschung, Planckstra\ss e 1, 64291 Darmstadt, Germany}
\affiliation{Theoretisch-Physikalisches Institut, Abbe Center of Photonics, \\ Friedrich-Schiller-Universit\"at Jena, Max-Wien-Platz 1, 07743 Jena, Germany}
\author{Fabian Sch\"utze}\email{fabian.schuetze@uni-jena.de}
\affiliation{Helmholtz-Institut Jena, Fr\"obelstieg 3, 07743 Jena, Germany}
\affiliation{GSI Helmholtzzentrum f\"ur Schwerionenforschung, Planckstra\ss e 1, 64291 Darmstadt, Germany}
\affiliation{Theoretisch-Physikalisches Institut, Abbe Center of Photonics, \\ Friedrich-Schiller-Universit\"at Jena, Max-Wien-Platz 1, 07743 Jena, Germany}

\date{\today}

\begin{abstract}
Studies of nonlinear quantum vacuum signals often model the driving laser fields as paraxial beams. This in particular holds for analytic approaches.
While this allows for reliable predictions in most situations, there are also notable exceptions. A prominent example is the overestimation of the polarization-flipped signal photon yield in the collision of two equally focused, parallel polarized laser beams by a factor of about six.
In the present work, we identify the origin of this deficiency and devise a strategy to obtain accurate closed-form expressions also in cases challenging the conventional (leading order) paraxial beam model.
We demonstrate the potential of our approach on the example of two linearly polarized laser pulses colliding at a generic collision angle.
\end{abstract}

\maketitle

\section{Introduction}
\label{sec:intro}

Quantum vacuum fluctuations give rise to effective nonlinear couplings between macroscopic electromagnetic fields \cite{Heisenberg:1935qt}.
So far, these nonlinearities could not be verified in a controlled laboratory experiment. However, recent progress in laser technology has brought their experimental detection in reach; see \cite{DiPiazza:2011tq,King:2015tba,Karbstein:2019oej,Fedotov:2022ely} for recent reviews.

All-optical signatures of quantum vacuum nonlinearity in laser beams collisions can be conveniently analyzed within the vacuum emission picture \cite{Galtsov:1971xm,Karbstein:2014fva}. This provides direct access to the directional emission characteristics and polarization properties of the signal far outside the interaction region of the laser fields.
To be experimentally accessible, the signal needs to differ in key-parameters, such as propagation direction, frequency or polarization, from the background of the driving {\it laser photons}.

In the context of the vacuum emission picture, the most advanced laser beam model numerically evolves the laser fields from an input configuration according to the linear Maxwell equations in vacuum \cite{Blinne:2018nbd}. For approaches aiming at numerically solving the fluctuation-induced nonlinear Maxwell equations in a general fashion, see \cite{Grismayer:2016cqp,Lindner:2021krv,Lindner:2022alm}.
On the other hand, analytical considerations typically model the laser beams as paraxial Gaussian beams supplemented with a temporal pulse envelope \cite{DiPiazza:2006pr,Tommasini:2010fb,King:2012aw,Gies:2017ygp}. 
It has been shown that this simplification allows for an accurate prediction of quantum vacuum signals in various experimentally relevant scenarios, like vacuum birefringence measurements \cite{Mosman:2021vua}.

However, by unveiling that it significantly overestimates the yield of polarization-flipped signal photons in the collision of two equally focused, parallel polarized laser beams, \cite{Blinne:2018nbd} identified a serious issue.
As will be demonstrated below, exactly the same problem arises for perpendicular polarized laser beams.
The present work aims to clarify the origin of this deficiency and to put forward a strategy allowing to overcome it while retaining the possibility of analytical insights.

For simplicity, here we limit our discussion to two linearly polarized laser beams colliding at an angle $\vartheta_{\rm coll}$ and only focus on the dominant polarization-flipped signal component, which originates in a quasi-elastic scattering process.
This signal is predominantly scattered in the vicinity of the forward cones of the driving laser beams for kinematic reasons. Inelastic scattering processes are typically suppressed relatively to elastic ones \cite{King:2015tba,Karbstein:2019oej}.
Because the effect is strongly suppressed for small $\vartheta_{\rm coll}$, a sizable quasi-elastic signal component is only generated for sufficiently different propagation directions, where the forward cones of the driving beams are well-separated. This immediately implies that the dominant signal component generically decomposes into two distinct contributions: a signal at the frequency of the first (second) beam induced in the effective interaction with the second (first) one. Its polarization-flipped component is polarized perpendicular to the first (second) beam.

Our article is organized as follows: after briefly recalling the underlying formalism and detailing the employed laser beam model in Sec.~\ref{sec:form}, we determine the directional emission characteristics of the polarization-flipped signal component and the associated integrated signal photon yield in Sec.~\ref{sec:results}.
Here, we derive compact analytical scalings for both of these quantities and confront them with numerical results. Finally, in Sec.~\ref{sec:concls} we end with conclusions and a outlook.

Throughout this work we use the metric convention $g^{\mu\nu}={\rm diag}(-1,1,1,1)$ and the Heaviside-Lorentz system with $c=\hbar=\varepsilon_0=1$. Correspondingly, the fine-structure constant is given by $\alpha=e^2/(4\pi)\simeq1/137$.

\section{Setup}
\label{sec:form}

\subsection{Formalism}

Vacuum fluctuations give rise to effective nonlinear couplings of electromagnetic fields beyond classical Maxwell theory ${\cal L}_{\rm MW}=-F_{\mu\nu}F^{\mu\nu}/4$ \cite{Heisenberg:1935qt,Weisskopf:1936hya,Schwinger:1951nm}.
For fields which are much weaker than the critical field $E_{\rm cr}=m^2/e\simeq1.3\times10^{18}\,{\rm V}/{\rm m}$ and vary on typical spatio-temporal scales $\lambda$ much larger than the Compton wavelength $\lambdabar_{\rm C}=1/m\simeq3.9\times10^{-7}\,\upmu{\rm m}$ of the electron, the leading effective interaction is quartic in the applied electromagnetic field and reads
\begin{align}
    {\cal L}_{\rm HE}^{1\text{-loop}}
    \simeq\frac{1}{1440\pi^2}\frac{e^4}{m^4}\Bigl[\bigl(F_{\mu\nu}F^{\mu\nu}\bigr)^2+\frac{7}{4}\bigl(F_{\mu\nu}{}^\star\!F^{\mu\nu}\bigr)^2\Bigr]\,. \label{eq:HElagrangian}
\end{align}
Corrections to \Eqref{eq:HElagrangian} are parametrically suppressed by powers of $\{\alpha,\,|eF^{\mu\nu}|/m^2,\,\lambdabar_{\rm C}/\lambda\}\ll1$. This criterion is perfectly fulfilled by the high-intensity laser fields currently available in the laboratory.

Photonic signatures of quantum vacuum nonlinearity induced by macroscopic electromagnetic fields can be conveniently analyzed within the vacuum emission picture \cite{Galtsov:1971xm,Karbstein:2014fva}, which is the approach we adopt here. It provides direct access to the angular emission characteristics of the signal photons far outside the interaction region of the driving laser beams.
The leading quantum vacuum signal is encoded in a zero-to-single signal photon emission process from the vacuum $|0\rangle$ to an out-state $\langle\gamma_p(\vec{k})|$ containing a single signal photon of wave vector $\vec{k}=|\vec{k}|(\cos\varphi\sin\vartheta,\sin\varphi\sin\vartheta,\cos\vartheta)$ and transverse polarization vector $\vec{\epsilon}_p(\vec{k})$; $\vartheta\in[0,\pi]$ and $\varphi\in[0,2\pi)$ are the polar and azimuthal angle, respectively.
The associated transition amplitude can be concisely expressed as \cite{Karbstein:2014fva}
\begin{equation}
    {\cal S}_p(\vec{k})\simeq\bigl\langle\gamma_p(\vec{k})\bigr|\int{\rm d}^4x\,\frac{\partial{\cal L}_{\rm HE}^{1\text{-loop}}}{\partial F^{\mu\nu}}(x)\,\hat{f}^{\mu\nu}(x)\bigl|0\bigr\rangle\,, \label{eq:amplitude}
\end{equation}
where $\hat{f}^{\mu\nu}(x)$ denotes the canonically quantized field strength tensor of the signal photon field.
The differential number of $p$-polarized signal photons of energy ${\rm k}=|\vec{k}|$ and emission direction $\vec{k}/{\rm k}$ then follows from \Eqref{eq:amplitude} as
\begin{equation}
    {\rm d}^3N_p(\vec{k})=\frac{{\rm d}^3k}{(2\pi)^3}\bigl|{\cal S}_p(\vec{k})\bigr|^2\,. \label{eq:diffphotonnumber}
\end{equation}

\subsection{Scenario}
\label{subsec:geomsetup}

In the present work, we consider the collision of two linearly polarized fundamental Gaussian paraxial laser beams $\ell\in\{1,2\}$ (wavelengths $\lambda_\ell=2\pi/\omega_\ell$) at a generic collision angle of $0\leq\vartheta_{\rm coll}\leq\pi$. The beam waists and Rayleigh ranges are $w_{0,\ell}$ and ${\rm z}_{\rm R,\ell}=\pi w_{0,\ell}^2/\lambda_\ell$, respectively. For convenience, we identify the beam axis $\vec{\kappa}_1=(0,0,1)$ of laser $\ell=1$ with the positive $\rm z$ axis, and the collision plane with the $\rm xz$ plane.
In turn, the beam axis of laser $\ell=2$ is directed along $\vec{\kappa}_2=(\sin\vartheta_{\rm coll},0,\cos\vartheta_{\rm coll})$.
The directions of the associated transverse vector potentials $\vec{A}_\ell(x)$ fulfilling $\vec{\kappa}_\ell\cdot\vec{A}_\ell(x)=0$ can then each be characterized by a single unit vector $\vec{a}_\ell$ \cite{Davis:1979,Barton:1989,Salamin:2007}. Here, we choose $\vec{a}_1=(\cos\beta_1,\sin\beta_1,0)$ and
$\vec{a}_2=(\cos\vartheta_{\rm coll}\cos\beta_2,\sin\beta_2,-\sin\vartheta_{\rm coll}\cos\beta_2)$ and parameterize these by the angles $\beta_\ell$.
Up to linear order in the expansion parameter $\theta_\ell=w_{0,\ell}/{\rm z}_{{\rm R},\ell}=2/(w_{0,\ell}\,\omega_l)$ of the paraxial approximation, the associated real electric and magnetic field vectors beam can be expressed as (our notations follow \cite{Salamin:2007}, but are adjusted to our conventions)
\begin{equation}
\begin{split}
 \vec{E}_\ell&=E_{0,\ell}\,{\rm e}^{-(\frac{r_\ell}{w_\ell})^2}\,\frac{w_{0,\ell}}{w_\ell}\Bigl(c_{1,\ell}\,\vec{a}_\ell+\theta_\ell\,\frac{{\rm x}_\ell}{w_\ell} s_{2,\ell}\,\vec{\kappa}_\ell\Bigr), \\
  \vec{B}_\ell&=E_{0,\ell}\,{\rm e}^{-(\frac{r_\ell}{w_\ell})^2}\,\frac{w_{0,\ell}}{w_\ell}\Bigl(c_{1,\ell}\,\vec{\kappa}_\ell\times\vec{a}_\ell+\theta_\ell\,\frac{{\rm y}_\ell}{w_\ell} s_{2,\ell}\,\vec{\kappa}_\ell\Bigr),
\end{split}
\label{eq:El,Bl}
\end{equation}
with beam radius $w_\ell=w_{0,\ell}\sqrt{1+({\rm z}_\ell/{\rm z}_{{\rm R},\ell})^2}$ and shorthand notations
\begin{equation}
\begin{split}
    c_{n,\ell}&=\cos(\psi_\ell-n\psi_{{\rm G},\ell})\,, \\
    s_{n,\ell}&=\sin(\psi_\ell-n\psi_{{\rm G},\ell})\,. 
\end{split}
\label{eq:cns/snl}
\end{equation}
The phases in \Eqref{eq:cns/snl} are defined as
\begin{equation}
    \begin{split}
    \psi_\ell&=\omega_\ell({\rm z}_\ell-t)+\frac{{\rm z}_\ell}{{\rm z}_{{\rm R},\ell}}\Bigl(\frac{r_\ell}{w_\ell}\Bigr)^2\,, \\
    \psi_{{\rm G},\ell}&=\arctan\Bigl(\frac{{\rm z}_\ell}{{\rm z}_{{\rm R},\ell}}\Bigr)\,,
\end{split}
\label{eq:phases}
\end{equation}
with $r_\ell=\sqrt{{\rm x}_\ell^2+{\rm y}_\ell^2}$.
To implement a finite pulse duration $\tau_\ell\gg1/\omega_\ell$, we moreover choose
\begin{equation}
    E_{0,\ell}={\cal E}_{0,\ell}\,{\rm e}^{-(\frac{{\rm z}_\ell-t}{\tau_\ell/2})^2}\,.
    \label{eq:envelope}
\end{equation}
The peak field amplitude ${\cal E}_{0,\ell}$ is then related to the laser pulse energy $W_\ell$ as \cite{Karbstein:2017jgh}
\begin{equation}
    {\cal E}_{0,\ell}^2\simeq8\sqrt{\frac{2}{\pi}}\frac{W_\ell}{\pi w_{0,\ell}^2\tau_\ell}\bigl[1+{\cal O}(\theta_\ell^2)\bigr]\,.
\end{equation}
Both beams reach their peak field in the focus at $\vec{x}=t=0$. 
Finally, the spatial components in Eqs.~\eqref{eq:El,Bl}-\eqref{eq:envelope} are
\begin{equation}
    {\rm x}_\ell=\vec{x}\cdot\vec{a}_{\ell}\,,\quad {\rm y}_\ell=\vec{x}\cdot(\vec{\kappa}_\ell\times\vec{a}_\ell)\,,\quad
    {\rm z}_\ell=\vec{x}\cdot\vec{\kappa}_\ell\,.
    \label{eq:xyz}
\end{equation}

Also note that $\theta_\ell$ amounts to the radial beam divergence.
In the limit of $\theta_\ell\to0$ \Eqref{eq:El,Bl} reduces to the (zeroth-order) Gaussian beam approximation that is widely and commonly employed in the study of all-optical quantum vacuum signatures.

Here, we limit ourselves to the signal component that scales linearly with the electromagnetic field (intensity) of laser $\ell=1$ on amplitude (probability) level and is
polarized perpendicular to the polarization state of this laser in its focus, i.e., fulfills $\vec{\epsilon}_p(\vec{k})\cdot\vec{E}_1\big|_{\vec{x}=0}=\vec{\epsilon}_p(\vec{k})\cdot\vec{a}_1=0$. We denote the associated polarization vector by
\begin{equation}
    \vec{\epsilon}_\perp(\vec{k})=\vec{e}_{(1)}\cos\beta_\perp+\vec{e}_{(2)}\sin\beta_\perp,
\end{equation}
with $\vec{e}_{(1)}=(\cos\varphi\cos\vartheta,\sin\varphi\cos\vartheta,-\sin\vartheta)$, $\vec{e}_{(1)}=(-\sin\varphi,\cos\varphi,0)$ and $\beta_\perp=\arctan\{\cot(\varphi-\beta_1)\cos\vartheta\}$.
Neglecting contributions of ${\cal O}(\theta_\ell^2)$ to the fields in \Eqref{eq:El,Bl} from the outset, the associated signal emission amplitude can be consistently determined therefrom only up to linear order in $\theta_\ell$.
Linearizing \Eqref{eq:amplitude} in the field components of laser $\ell=1$ and omitting contributions that scale manifestly quadratic in $\theta_2$, we obtain
\begin{align}
    {\cal S}_\perp(\vec{k})\simeq\,&\frac{\rm i}{\pi}\frac{m^2}{45}\sqrt{\frac{\alpha}{\pi}\frac{\rm k}{2}}\Bigl(\frac{e}{m^2}\Bigr)^3 \nonumber\\
    &\times\Bigl\{\cos\beta_\perp\bigl[\vec{e}_{(1)}\cdot\vec{U}(\vec{k})-\vec{e}_{(2)}\cdot\vec{V}(\vec{k})\bigr] \nonumber\\
    &\quad+\sin\beta_\perp\bigl[\vec{e}_{(1)}\cdot\vec{V}(\vec{k})+\vec{e}_{(2)}\cdot\vec{U}(\vec{k})\bigr]\Bigr\}\,.
    \label{eq:Sperp}
\end{align}
Here, we use the following convention for the on-shell Fourier transform from position to momentum space,
\begin{equation}
    \vec{U}(\vec{k})=\int{\rm d}^4x\,{\rm e}^{-{\rm i}(\vec{k}\cdot\vec{x}-{\rm k}t)}\,\vec{U}(x)\,,
    \label{eq:U}
\end{equation}
with the vectors in position space given by
\begin{equation}
\begin{split}
    \vec{U}(x)&=4{\cal F}_{12}\vec{E}_2-7{\cal G}_{12}\vec{B}_2\,, \\
    \vec{V}(x)&=4{\cal F}_{12}\vec{B}_2+7{\cal G}_{12}\vec{E}_2\,,
\end{split}
\label{eq:U,V}
\end{equation}
where we introduced the shorthand notations
\begin{equation}
    \begin{split}
    {\cal F}_{12}&=\frac{1}{2}(\vec{B}_1\vec{B}_2-\vec{E}_1\vec{E}_2)\quad\text{and}\\
    {\cal G}_{12}&=-\frac{1}{2}(\vec{B}_1\vec{E}_2+\vec{E}_1\vec{B}_2)\,.
\end{split}
\label{eq:F12,G12}
\end{equation}
The signal arising from \Eqref{eq:Sperp} can be interpreted in terms of $\ell=1$ laser photons that are quasi-elastically scattered into a $\perp$-polarized mode via the effective interaction with laser $\ell=2$.
For kinematic reasons these are predominantly scattered in the direction of $\vec{\kappa}_1$ which amounts to signal photon wave vectors with $\vartheta\ll1$.
The analogous signal polarized perpendicular to laser $\ell=2$ can be effectively inferred therefrom by appropriately relabeling the beam parameters and mapping the electromagnetic fields with $\ell=1$ on those with $\ell=2$ and vice versa.
We emphasize that the only remaining nontrivial task in evaluating the signal photon amplitude~\eqref{eq:Sperp} is to perform the Fourier integrals.

\subsection{Simplifying assumptions}
\label{subsec:beams}

Equations~\eqref{eq:El,Bl}-\eqref{eq:xyz} imply that a laser $\ell$ can reach a substantial fraction of its peak field only within a cylinder $V_\ell$ of radius $w_{0,\ell}$ and length set by ${\rm z}_{{\rm R},\ell}$ about its focus at $\vec{x}=0$. Away from the focus, the beam radius $w_\ell$ increases with the longitudinal beam coordinate $\rm z$ and, correspondingly, the field strength drops.
For $\tau_\ell\gtrsim2{\rm z}_{{\rm R},\ell}$ the field may be strong within the whole cylinder at a given time, while for $\tau_\ell\lesssim2{\rm z}_{{\rm R},\ell}$ the strong field is limited to a segment that travels along $\rm z$ with time.
Currently available tightly focused near-infrared high-intensity lasers relevant for the study of quantum vacuum signals typically fulfill $\tau_\ell\gtrsim2{\rm z}_{{\rm R},\ell}$.
Recall, that $w_{0,\ell}$ is the $1/{\rm e}^2$ focus radius on intensity level, and ${\rm z}_{{\rm R},\ell}$ measures the distance from focus over which the on-axis intensity drops by a factor of $1/2$.
To put both scales on the same footing, we therefore introduce the cylinder's half-length as $l_\ell={\rm z}_{{\rm R},\ell}\sqrt{{\rm e}^2-1}$ and define $V_\ell=2l_\ell(\pi w_{0,\ell}^2)$.
For state-of-the-art laser parameters sizable quantum vacuum signals are only induced in the interaction region of the colliding laser fields, which amounts to the intersection of $V_1$ and $V_2$ in the present scenario.

\begin{figure}
  \centering
  \includegraphics[width=0.8\linewidth]{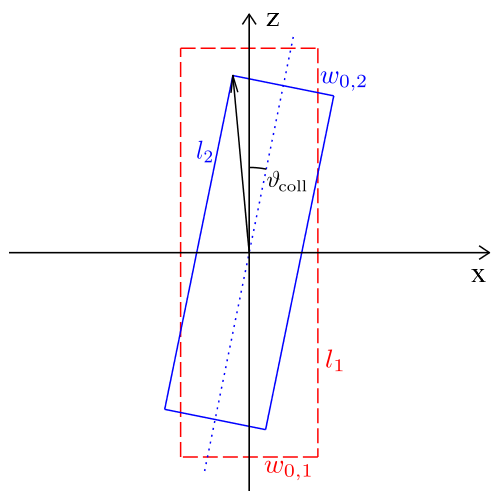}
\caption{Projection of the strong field volumes $V_1$ (blue solid) and $V_2$ (red dashed) onto the collision plane. The intersection $V_1\cap V_2$ of these volumes becomes independent of $l_2\leq l_1$ once the top left corner of $V_2$ reaches the right edge of $V_1$.}
\label{fig:inf_Ray_cond}
\end{figure}

Accounting for the fact that for $\theta_\ell\ll1$ we have $w_{0,\ell}\ll l_\ell$, the above considerations imply that typically there exists a range of collision angles for which the interaction region $V_1\cap V_2$ becomes independent of the length scales $l_\ell$, and thus the Rayleigh ranges ${\rm z}_{{\rm R},\ell}$ \cite{Karbstein:2021dz}.
See Fig.~\ref{fig:inf_Ray_cond} for an illustration.
The range of collision angles for which this is the case can be inferred from elementary geometric considerations.
Considering the projection of $V_1$ and $V_2$ on the collision plane, one can easily establish that, w.l.o.g. presuming $l_1\geq l_2>w_{0,1}$, this holds true for
\begin{equation}\label{eq:IRLAcollangle}
     \sin\vartheta_{\rm coll} \gtrsim \frac{\frac{w_{0,1}}{l_{2}} + \frac{w_{0,2}}{l_{2}}\sqrt{1-(\frac{w_{0,1}}{l_{2}})^2+(\frac{w_{0,2}}{l_{2}})^2}}{1+(\frac{w_{0,2}}{l_{2}})^2}\,.
\end{equation}
For $w_{0,1}=w_{0,2}=w_0$ this expression simplifies to
\begin{equation}
     \sin\vartheta_{\rm coll} \gtrsim \frac{2\frac{w_0}{l_2}}{1+(\frac{w_0}{l_2})^2}\, \xrightarrow{w_0\ll l_2}\,2\frac{w_0}{l_2}\,.
     \label{eq:IRLAcollangle1}
\end{equation}
Hence, specifically for two lasers of the same wavelength $\lambda=800\,{\rm nm}$ focused to $w_0=2\lambda$ as considered in Sec.~\ref{sec:results} the above criterion if fulfilled for collision angles in the range  $7.2^\circ \lesssim \vartheta_{\rm coll} \lesssim 172.8^\circ$.

It has been argued that in the parameter regime where the interaction region becomes independent of the Rayleigh ranges of the colliding beams it amounts to a good approximation to formally send ${\rm z}_{{\rm R},\ell}\to\infty$ while keeping $w_{0,\ell}$ and $\lambda_\ell$ fixed when determining the quantum vacuum signal \cite{Gies:2017ygp,King:2018wtn}. This approximation is often referred to as {\it infinite Rayleigh range/length approximation}. Previous studies limited themselves to \Eqref{eq:El,Bl} at zeroth order in the paraxial approximation, i.e., only kept contributions of ${\cal O}(\theta_\ell^0)$.
In the present work, we emphasize that whereas taking the above limit for the field amplitude profiles indeed yields a good approximation, this is not necessarily the case for the directional properties of the fields. In order to reproduce the correct behavior of the polarization-flipped signal for the case of parallel or perpendicular polarized lasers, one is required to also keep the expansion parameter $\theta_\ell$ in \Eqref{eq:El,Bl} finite; see the explicit examples given below.
Hence, we advocate amending the definition of the infinite Rayleigh range/length approximation as follows: take the formal limit of ${\rm z}_{{\rm R},\ell}\to\infty$ in the field amplitude profiles and keep $w_{0,\ell}$, $\lambda_\ell$ and $\theta_\ell$ (at the relevant order of the paraxial expansion; see below) finite.
Implementing this approximation for Eqs.~\eqref{eq:El,Bl}-\eqref{eq:phases}, we obtain
\begin{equation}
\begin{split}
 \vec{E}_\ell&=E_{0,\ell}\,{\rm e}^{-(\frac{r_\ell}{w_{0,\ell}})^2}\,\Bigl(c_{1,\ell}\,\vec{a}_\ell+\theta_\ell\,\frac{{\rm x}_\ell}{w_{0,\ell}} s_{2,\ell}\,\vec{\kappa}_\ell\Bigr), \\
  \vec{B}_\ell&=E_{0,\ell}\,{\rm e}^{-(\frac{r_\ell}{w_{0,\ell}})^2}\,\Bigl(c_{1,\ell}\,\vec{\kappa}_\ell\times\vec{a}_\ell+\theta_\ell\,\frac{{\rm y}_\ell}{w_{0,\ell}} s_{2,\ell}\,\vec{\kappa}_\ell\Bigr),
\end{split}
\label{eq:El,Bl_IRRA}
\end{equation}
with
\begin{equation}
\begin{split}
    c_{n,\ell}&=\cos\bigl(\omega_\ell({\rm z}_\ell-t)\bigr)\,, \\
    s_{n,\ell}&=\sin\bigl(\omega_\ell({\rm z}_\ell-t)\bigr)\,. 
\end{split}
\label{eq:cns/snl_IRRA}
\end{equation}
From these expressions it is obvious that, adopting the simplifying assumptions just detailed, all the Fourier integrations in Eqs.~\eqref{eq:Sperp}-\eqref{eq:F12,G12} that need to be performed when determining the signal photon amplitude up to ${\cal O}(\theta)$, with $\theta\sim\theta_1\sim\theta_2$, are reduced to elementary Gaussian integrals that can be readily carried out analytically.
As obvious from \Eqref{eq:U}, any powers of $\rm x$, $\rm y$ and $\rm z$ multiplying the exponential function can be dealt with by parameter differentiations for $k_{\rm x}$, $k_{\rm y}$ and $k_{\rm z}$, respectively. 
We also remark that it is clear that the present approach can be readily generalized to higher orders in the paraxial expansion.
Actually performing the four fold integral is somewhat tedious and results in a rather unhandy expression for the signal photon amplitude. For this reason we do not reproduce it here.
However, note that the corresponding expression for the contribution at ${\cal O}(\theta^0)$ can be found in \cite{Karbstein:2021dz}.
In order to isolate the most relevant physical parameter dependencies and scalings, and also with regard to the subsequent determination of the associated signal photons via \eqref{eq:diffphotonnumber},
additional simplifications are therefore highly desirable.

To achieve this, we employ the approximation strategy devised in \cite{Karbstein:2018omb,Karbstein:2021dz}, which makes use of the fact that the dominant signal arises from the quasi-elastic scattering of laser $\ell=1$ photons and thus is characterized by an energy of ${\rm k}\simeq\omega_1$. Correspondingly, $\delta\hat{\rm k}=({\rm k}-\omega_1)/\omega_1\ll1$ amounts to a small parameter.
For kinematic reasons, the associated signal photons are then emitted in directions close to the forward beam axis of laser $\ell=1$ and  fulfill $\vartheta\ll1$, constituting a second small parameter.
For simplicity, we moreover limit our explicit considerations to laser pulses of the same pulse duration $\tau=\tau_1=\tau_2$.

Upon performing the Fourier integration~\eqref{eq:U} over the position-space vectors introduced in \Eqref{eq:U,V}, \Eqref{eq:Sperp} decomposes into a sum of terms. An explicit restriction to the quasi-elastic scattering signal is equivalent to keeping only those contributions that do not depend on the oscillation frequency $\omega_2$ of laser $\ell=2$ \cite{Karbstein:2019oej}.
This $\omega_2$ independence of the signal implicates that in the evaluation of \Eqref{eq:Sperp} the quadratic dependencies on the electromagnetic fields~\eqref{eq:El,Bl_IRRA} of laser $\ell=2$ are effectively replaced by their cycle-averages. 
Upon recasting factors of $\rm x$, $\rm y$ and $\rm z$ multiplying the exponential functions by parameter differentiations for the associated momentum components, all Fourier integrals~\eqref{eq:U} to be performed are over Gaussian functions in coordinate space. 
As the Fourier transform of a Gaussian yields again a Gaussian, we have thus established that the signal photon amplitude is quadratic in all momentum components.
Besides, it is easy to establish that the resulting expression for \Eqref{eq:Sperp} is characterized by an overall factor of
\begin{equation}
{\cal S}_\perp(\vec{k})\sim{\rm e}^{-\frac{1}{3}(\frac{\tau\omega_1}{4})^2\delta\hat{\rm k}^2}\,.
\label{eq:Sperp_scaling}
\end{equation}
This dependency manifestly ensures that for $\tau\omega_1\gg1$, as considered throughout this work, non-vanishing contributions to the signal are indeed limited to $\delta\hat{\rm k}\ll1$; the typical decay width is set by the dimensionless parameter $1/(\tau\omega_1)$.
One can easily verify that the other dependencies on $\delta\hat{\rm k}$ in the exponential of ${\cal S}_\perp(\vec{k})$ scale as $\vartheta^2\delta\hat{\rm k}$ and $(\vartheta\delta\hat{\rm k})^2$ for $\vartheta^2\ll1$.
Given that $w_0\sim w_{0,1}\sim w_{0,2}$, the decay of the signal with $\vartheta$ is governed by a divergence proportional to $\theta\sim1/(w_0\omega_1)$.
Hence, one can safely neglect these additional dependencies as long as $\tau\gg w_0$, which is precisely the parameter regime considered here.
Moreover, all contributions $\sim(\delta\hat{\rm k})^n$ with $n\in\mathbb{N}$ in the prefactor of the exponential are then subleading in comparison to the term $\sim(\delta\hat{\rm k})^0$ and thus can also be neglected.
In summary, this approximation amounts to identifying ${\rm k}=\omega_1$ everywhere in ${\cal S}_\perp(\vec{k})$ aside from the factor in \eqref{eq:Sperp_scaling}.
Plugging the resulting approximation for ${\cal S}_\perp(\vec{k})$ into \Eqref{eq:diffphotonnumber}, the integration over signal photon energy can be performed by identifying ${\rm k}^2\,{\rm dk}\,\to\,\omega_1^2\,{\rm dk}$ and formally extending the integration domain to $-\infty\leq{\rm k}\leq\infty$. One can easily check that in the considered parameter regime this is perfectly justified. 
The other approximation to be invoked is an explicit restriction to the leading contributions to ${\cal S}_\perp(\vec{k})$ for $\vartheta\ll1$. Here, we keep the leading contribution $\sim\vartheta^2$ in the argument of the exponential function, and similarly expand the terms in its prefactor up to quadratic order in $\vartheta$.
When integrating the modulus square of the signal emission amplitude over the polar angle $\vartheta$ in a later step, we moreover approximate $\int_{-1}^1{\rm d}\!\cos\vartheta\,\to\int_0^\infty{\rm d}\vartheta\,\vartheta$ consistent with the expansion for $\vartheta\ll1$ just outlined.

Finally, we highlight the following, somewhat subtle but very important point: The Gaussian nature of ${\cal S}_\perp(\vec{k})$ prior to implementing the parameter differentiations noted above \Eqref{eq:Sperp_scaling} implies that in the most general situation the parameter differentiation for a given momentum component may result in terms linear in this as well as all other momentum components. Accounting for the fact that for $\vartheta\ll1$ we have $k_{\rm x}\sim k_{\rm y}\sim \omega_1\vartheta$ and $k_{\rm z}\sim\omega_1$, and that the leading $\vartheta$ dependent contribution in the argument of the exponential function scales quadratic with $\vartheta$, we can immediately exclude the existence of terms scaling as $k_{\rm x}k_{\rm z}\sim k_{\rm y}k_{\rm z}\sim k_{\rm x}{\rm k}\sim k_{\rm y}{\rm k}\sim\vartheta$ in the argument of the exponential; these receive a parametric suppression with additional powers of $\vartheta$. From this, we know that factors of $\rm x$ and $\rm y$ in the integrand of ${\cal S}_\perp(\vec{k})$ translate into factors scaling as $\partial_{k_{\rm x}}\sim\partial_{k_{\rm y}}\sim k_{\rm x}\sim k_{\rm y}\sim\omega_1\vartheta$ in momentum space. At the same time, the fact that with the above approximations the leading dependence on $k_{\rm z}^2\sim k_{\rm z}{\rm k}\sim{\rm k}^2$ is encoded in the overall exponential factor in \Eqref{eq:Sperp_scaling} implies that factors of $\rm z$ translate into factors $\sim\partial_{k_{\rm z}}\sim\delta\hat{\rm k}$ to be set identically to zero within the approximation strategy devised above.

These considerations allow us to infer that a given contribution to the signal emission amplitude at $n$th order in the paraxial approximation is generically suppressed by an overall factor of $\theta^n$ if its prefactor does not involve factors of ${\rm x}$, $\rm y$ and $\rm z$. On the other hand, if a contribution at ${\cal O}(\theta^n)$ comes with a finite power $j$ of ${\rm x}\sim{\rm y}$ it scales at most as $\theta^n(\omega_1\vartheta)^j$. Because of $\vartheta\lesssim\theta\sim1/\omega_1$ this again translates into a factor of $\theta^n$. A term multiplied by powers of $\rm z$ can be neglected from the outset. Hence, we have explicitly shown that higher-order contributions in the paraxial approximation indeed receive a parametric suppression with powers of $\theta$ also on the level of the signal emission amplitude. 
As different orders $n$, $n'$ in the paraxial approximation may in addition receive parametric suppressions with distinct powers of $\vartheta\ll1$, terms with $n\neq n'$ may in effect still exhibit the same scaling with $\theta$: For instance a contribution to the signal emission amplitude proportional to $\theta^0\vartheta$ effectively contributes at the same order as the one $\sim\theta(\omega_1\vartheta)$. Both of these effectively scale linear with $\theta$.

\section{Results}\label{sec:results}

Here, we adopt the simplifying assumptions and approximations detailed in Sec.~\ref{subsec:beams}, and use the shorthand notation
\begin{equation}
\begin{split}
    &b(x) = 4x(1-\cos\vartheta_{\rm coll})^2+\Bigl(\frac{\tau}{\bar w_0}\Bigr)^2 \sin^2\vartheta_{\rm coll}\,,\\
    &\text{with}\quad \bar{w}_0^2=\frac{2w_{0,1}^2+w_{0,2}^2}{3}\,.
\end{split}
\end{equation}
Inserting the modulus square of the appropriately simplified signal emission amplitude~\eqref{eq:Sperp} accurate up to linear order in $\theta$ into \eqref{eq:diffphotonnumber},
we arrive at the following result for the directional emission characteristics of the signal photons,
\begin{equation}\label{eq:photondensity2ndorder}
\begin{split}
    & \frac{{\rm d}^2 N_\perp}{{\rm d}\varphi\,{\rm d}\!\cos\vartheta} \approx \frac{8\sqrt{3}\alpha^4}{2025\pi^3}\frac{W_1 W_2^2 \omega_1^3}{m^8 w_{0,1}^2} \Bigl(\frac{w_{0,1}}{\bar w_0}\Bigr)^4  \\
    & \times \frac{( 1-\cos\vartheta_{\rm coll})^2}{\sqrt{b(1) b(3)}}\bigl( n_{\perp}^{(0)} + n_{\perp}^{(1)} \vartheta +  n_{\perp}^{(2)}\vartheta^2 \bigr) \\
    & \times \exp \biggl\{\!-\frac{1}{3}\Bigl(\frac{w_{0,1}}{\bar w_0}\omega_1\vartheta\Bigr)^2\Bigl[w_{0,1}^2 \frac{b(0) }{b(3)}\cos^2\varphi+\frac{w_{0,2}^2}{2}\Bigr]\!\biggr\},
\end{split}
\end{equation}
where
\begin{equation}
    n_{\perp}^{(0)} = ( 1-\cos\vartheta_{\rm coll})^2 \sin^2[2(\beta_1+\beta_2)]\,.
\end{equation}
The explicit expressions for the coefficients $n_{\perp}^{(n)}$  of the contributions $\sim\vartheta^n$ with $n\in\{1,2\}$ multiplying the exponential in \Eqref{eq:photondensity2ndorder} are given in Appendix~\ref{sec:nperp}.
We emphasize that in order to make transparent which terms originate from contributions at zeroth and linear order in the paraxial expansion of the laser fields, throughout this work we do not express any other parameters, like the product $w_{0,\ell}\omega_\ell$, in terms of $\theta_\ell$. This allows, for instance, to recover the result at zeroth order in the paraxial expansion by simply setting $\theta_\ell=0$.
Noteworthy, \Eqref{eq:Sperp} does not give rise to contributions linear in $\theta_2$, such that \Eqref{eq:photondensity2ndorder} depends only on $\theta_1$. 
Also note that upon integrating \Eqref{eq:photondensity2ndorder} over the full azimuthal angle $\varphi$, contributions odd in $\vartheta$ vanish.
In the remainder of this work, we refer to all results determined using \Eqref{eq:photondensity2ndorder} with $\theta_1=0$ ($\theta_1\neq0$) as {\it (next-to-)leading-order} or (N)LO in the paraxial approximation.

Some clarifications are in order here:
As we neglected terms of ${\cal O}(\theta^2)$ to the signal emission amplitude, for $n_\perp^{(0)}\neq0$ only contributions to \Eqref{eq:photondensity2ndorder} up to linear order in $\theta$ can be considered as consistent. Hence, when employing \Eqref{eq:photondensity2ndorder} for this case we set the term scaling quadratic with $\theta_1$ in $n_\perp^{(2)}$ to zero. Higher-order corrections are relatively suppressed by factors of $\theta\ll1$. Note that for all physically relevant scenarios fulfilling $\vartheta_{\rm coll}\neq0$ the condition $n_\perp^{(0)}\neq0$ holds true for all laser polarization choices except from $\tilde\beta=2(\beta_1+\beta_2)=j\pi$ with $j\in\mathbb{Z}$.
On the other hand, for $\tilde\beta=j\pi$ both $n_\perp^{(0)}$ and $n_\perp^{(1)}$ vanish identically, which implies that the leading signal is parametrically suppressed by a factor of $\vartheta^2\ll1$ relative to the $n_\perp^{(0)}\neq0$ case.
The contributions to the signal emission amplitude then scale as ${\cal S}_\perp(\vec{k})\sim\vartheta(c_0\theta^0+c_1\theta\omega_1)$, with coefficients $c_0$ and $c_1$. Because of $\theta\omega_1\sim\theta^0$ and the fact that the omitted ${\cal O}(\theta^2)$ terms to ${\cal S}_\perp(\vec{k})$ clearly cannot contribute to the differential signal photon number at ${\cal O}(\theta^2)$, in this case all terms in \Eqref{eq:photondensity2ndorder} are to be accounted for; see also the last paragraph of Sec.~\ref{subsec:beams}. The neglected subleading corrections are parametrically suppressed by at least a factor of $\theta\ll1$.

Integrating \Eqref{eq:photondensity2ndorder} over the angles $\varphi$ and $\vartheta$, we obtain
\begin{equation}
\begin{aligned}
    N_\perp\approx\ & \frac{16\alpha^4}{675\pi^2}\frac{W_1W_2^2\omega_1}{m^8\bar w_0^3w_{0,2}} \frac{(1-\cos\vartheta_{\rm coll})^2}{\sqrt{b(1)b\bigl((\frac{w_{0,2}}{\bar w_0})^2\bigr)}} \\
    & \times \biggl[n^{(0)}_{\perp} + \Bigl(\frac{1}{w_{0,1}\omega_1}\frac{\bar w_0}{w_{0,2}}\Bigr)^2\frac{b(3)}{b\bigl((\frac{w_{0,2}}{\bar w_0})^2\bigr)}N^{(2)}_{\perp}\biggr],    
\end{aligned}\label{eq:Nperp}
\end{equation}
with the expression for $N_\perp^{(2)}$ given in Appendix~\ref{sec:Nperp}.
Subsequently, we compare and benchmark our analytic result~\eqref{eq:Nperp} with the outcome of {\it fully numerical} calculations modeling the laser beams as solutions of the linear Maxwell equations in vacuum.
To be specific, in the numerical vacuum emission solver~\cite{Blinne:2018nbd} we initialize the colliding laser fields by the complex analogues of the real-valued electric fields in \Eqref{eq:El,Bl} at $t=0$ and $\theta_\ell=0$, i.e., at zeroth order in the paraxial approximation.
In addition, we numerically evaluate \Eqref{eq:Sperp} for the fields in \Eqref{eq:El,Bl} with $\theta_\ell=0$ and determine the associated signal photon number at leading order, i.e., {\it LO numerical}, via \Eqref{eq:diffphotonnumber}.

In our explicit examples, we assume the strong laser fields to be provided by two identical state-of-the-art petawatt-class lasers delivering pulses of energy $W_\ell=25\,{\rm J}$ and duration $\tau = 25\,{\rm fs}$ at a central wavelength of $\lambda_\ell=800\,{\rm nm}$. 
For the waist sizes we choose the experimentally feasible value of $w_{0,\ell}=2\lambda_\ell=1.6\,\upmu{\rm m}$.
This fixes all parameters aside from $\vartheta_{\rm coll}$ and $\beta_\ell$ in Eqs.~\eqref{eq:photondensity2ndorder} and \eqref{eq:Nperp}; note that in this case we have $\bar w_0=w_{0,1}=w_{0,2}$.

\begin{figure}
  \centering
  \includegraphics[width=0.9\linewidth]{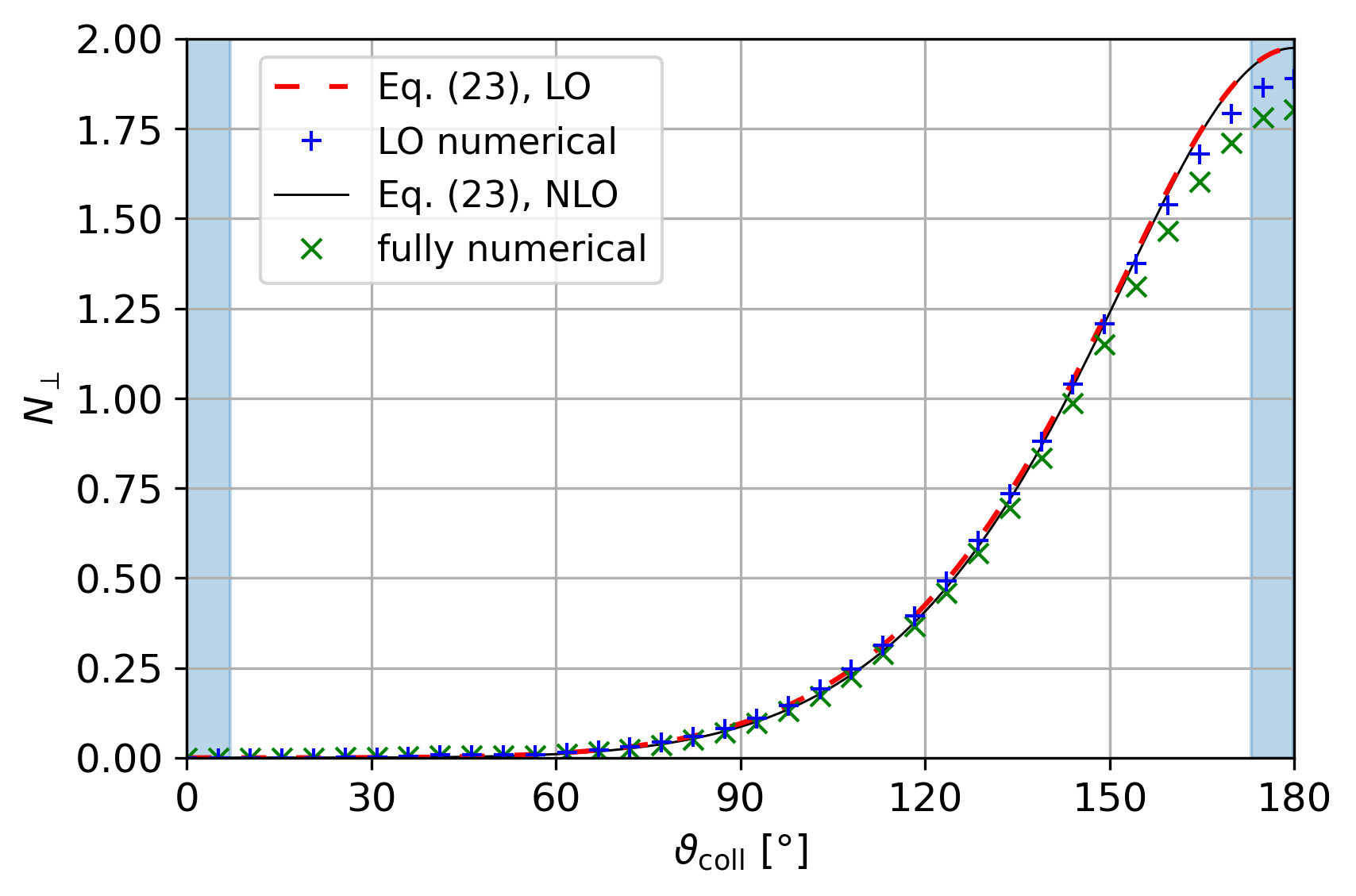}
\caption{Results for $N_\perp$ as a function of $\vartheta_{\rm coll}$ for $\beta_1=2\beta_2=\pi/2$. The values of $\vartheta_{\rm coll}$ outside the regime of validity of our analytic approximation~\eqref{eq:IRLAcollangle1} are shaded in blue.
The green crossed ($\times$) data points are obtained from a fully numerical calculation modeling the laser beams as solutions of the linear Maxwell equations. The blue crossed ($+$) data points are for the collision of two pulsed Gaussian beams at zeroth order in the paraxial approximation. The red dashed and black solid lines are plots of \Eqref{eq:Nperp} at LO and NLO.}
\label{fig:Nperp45}
\end{figure}
Figure~\ref{fig:Nperp45} shows results for $N_\perp$ as a function of $\vartheta_{\rm coll}$ for $\beta_1=2\beta_2=\pi/2$. This choice, enforcing a relative polarization angle of $\pi/4$ between the two laser fields colliding in the focus, is known to maximize the $\perp$-polarized signal component for generic $\vartheta_{\rm coll}$ \cite{Karbstein:2021dz}. Here, we plot the curves obtained by evaluating \Eqref{eq:Nperp} at LO and NLO together with the outcomes of LO and fully numerical calculations.
In line with \Eqref{eq:IRLAcollangle1}, constraining the range of applicability of our analytical approximations, for $\vartheta_{\rm coll} \lesssim 170^\circ$ the four curves displayed agree reasonably well.
The small deviation between the analytical and numerical results in this collision angle regime can be traced back to the employed approximation schemes underlying the analytical calculations detailed in Sec.~\ref{subsec:beams}. As is well-known, for this choice of the beam polarizations the maximal signal is obtained in a counter-propagating geometry \cite{Bialynicka-Birula:1970nlh,Brezin:1971nd}.

\begin{figure}
  \centering
  \includegraphics[width=0.9\linewidth]{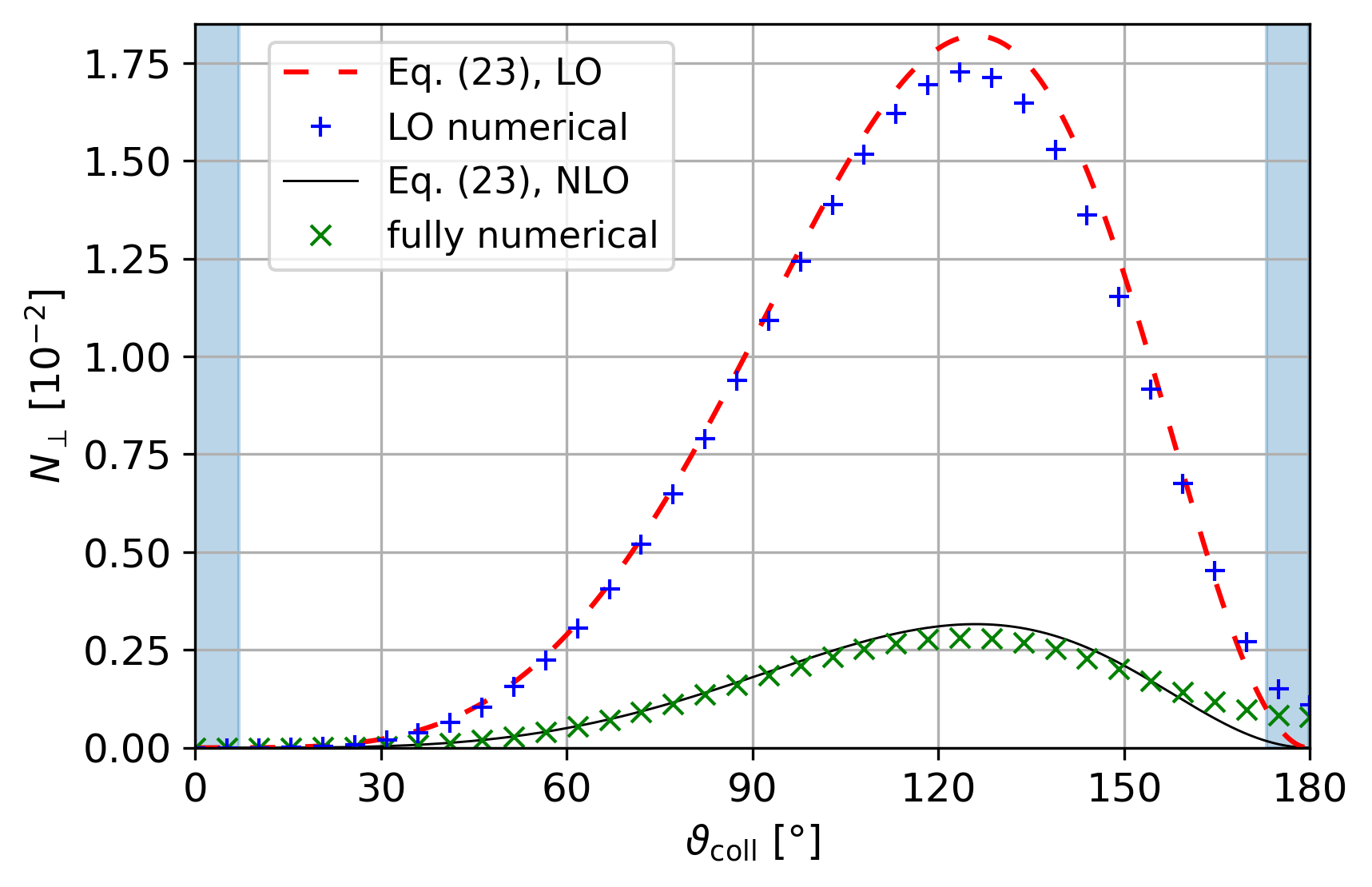}
\caption{Results for $N_\perp$ as a function of $\vartheta_{\rm coll}$ for $\beta_1=\beta_2=\pi/2$. The values of $\vartheta_{\rm coll}$ outside the regime of validity of our analytic approximation~\eqref{eq:IRLAcollangle1} are shaded in blue. The green crossed ($\times$) data points are obtained from a fully numerical calculation modeling the laser beams as solutions of the linear Maxwell equations. The blue crossed ($+$) data points are for the collision of two pulsed Gaussian beams at zeroth order in the paraxial approximation. The red dashed and black solid lines are plots of \Eqref{eq:Nperp} at LO and NLO.}
\label{fig:Nperp0}
\end{figure}
Figure~\ref{fig:Nperp0} displays similar results as those in Fig.~\ref{fig:Nperp45}, but for a different orientation of the polarization vectors of the colliding laser beams. Now the laser fields fulfill $\beta_1=\beta_2=\pi/2$ and thus are parallel polarized in the focus.
Interestingly, for this choice of the polarization vectors the leading order paraxial results for $N_\perp$, obtained by evaluating \Eqref{eq:Nperp} at LO and determined from an LO numerical calculation, differ significantly from those going beyond, namely \Eqref{eq:Nperp} at NLO and the outcome of a fully numerical calculation. The former considerably overestimate the physical value of $N_\perp$ following from the latter in a wide range of collision angles.
This pronounced discrepancy was first noticed in \cite{Blinne:2018nbd} for the collision of two laser beams focused to waist sizes equal to the laser wavelength.
By simply dividing the results for $N_\perp$ inferred from \Eqref{eq:Nperp} at LO and NLO, we can now even infer an analytic estimate for the discrepancy between the LO and NLO results depicted in Fig.~\ref{fig:Nperp0}. The resulting expression for general beam waists is independent of $\vartheta_{\rm coll}$ and reads
\begin{equation}
    \frac{N_\perp}{N_\perp|_{\theta_1=0}}=1-\frac{7}{6}\Bigl(\frac{w_{0,2}}{\bar w_0}\Bigr)^2\biggl[1-\frac{7}{24}\Bigl(\frac{w_{0,2}}{\bar w_0}\Bigr)^2\biggr]\,. \label{eq:Nratio}
\end{equation}
Equation~\eqref{eq:Nratio} depends only on the ratio $w_{0,2}/\bar w_0$.
Accounting for $0\leq(w_{0,2}/\bar w_0)^2\leq3$, it is then easy to show that $N_\perp \leq N_\perp|_{\theta_1=0}$. For the specific example of $\bar w_0=w_{0,1}=w_{0,2}$ considered here, \Eqref{eq:Nratio} yields $N_\perp/N_\perp|_{\theta_1=0} = 25/144 \approx 1/6$.
All curves displayed in Fig.~\ref{fig:Nperp0} exhibit a distinct maximum $N_\perp|_{\rm max}$ located at $0^\circ<\vartheta_{\rm coll}|_{\rm max}<180^\circ$.
Equation~\eqref{eq:Nperp} predicts the maximum value at both LO and NLO to be achieved for 
\begin{equation}\label{eq:optcollangle}
    \cos\vartheta_{\rm coll}|_{\rm max} = \frac{2+(\frac{\tau}{w_0})^2 - 2\sqrt{1+2(\frac{\tau}{w_0})^2}}{4-(\frac{\tau}{w_0})^2}\,,
\end{equation}
i.e., to be fully determined by the pulse durations and the waist sizes of the colliding laser fields. See Tab. \ref{tab:theta_max} for the specific values.
\begin{table}[]
    \centering
    \begin{tabular}{l|c|c}
        & $N_{\perp}|_{\rm max}\,[10^{-2}]$ & $\vartheta_{\rm coll}|_{\rm max}$\\
        \hline \hline
        \Eqref{eq:Nperp}, LO & $1.82$ & $126.0^\circ$  \\
         LO numerical & $1.73$ & $124.6^\circ$  \\
        \Eqref{eq:Nperp}, NLO & $0.32$ & $126.0^\circ$  \\
         fully numerical & $0.29$ & $123.9^\circ$ 
    \end{tabular}\caption{Maximum values of $N_\perp$ and associated collision angles in Fig.~\ref{fig:Nperp0}. $N_\perp|_{\rm max}$ is significantly reduced when increasing the order of the paraxial approximation from LO to NLO.}
    \label{tab:theta_max}
\end{table}
The four calculations performed predict the maximum to be reached for  $\vartheta_{\rm coll}\approx124^\circ\ldots126^\circ$.
We emphasize that the analytical predictions of \Eqref{eq:Nratio} based on several simplifying assumptions and approximations (recall Sec.~\ref{subsec:beams}) are in reasonably good agreement with the corresponding numerical calculations.
Especially with regard to the NLO prediction of \Eqref{eq:Nratio} we recall that here it is compared with a fully numerical calculation solving the linear Maxwell equations; the relative deviation is $\lesssim10\%$. The latter thus effectively accounts for arbitrarily high orders of the paraxial approximation.
In the explicit example considered here the expansion parameter governing the paraxial approximation is $\theta_1=\theta_2=1/(2\pi)\approx0.16$. This value both hints at the consistency of our findings and suggests that in order to achieve a relative deviation below the ${\cal O}(10)\%$ level for the full range of collision angles $\vartheta_{\rm coll}$ in \Eqref{eq:IRLAcollangle1} and all possible laser polarizations $\beta_\ell$ higher-order corrections in the paraxial and $\vartheta$ expansions need to be included in our analytical approximation.
For completeness, we also note that we have explicitly checked that the value of $N_\perp=0$ obtained from \Eqref{eq:Nperp} for this choice of the laser beam polarizations and $\vartheta_{\rm coll}=180^\circ$, which is outside the regime of applicability of our analytical approximation constrained by \Eqref{eq:IRLAcollangle1}, is promoted to a nonzero value at higher-orders in the $\vartheta$ expansion.

In the next step, we provide a set of simple physical arguments hinting at why the LO paraxial approximation fails in the quantitatively accurate prediction of the $\perp$-polarized signal for the cases where $\tilde\beta=j\pi$ with $j\in\mathbb{Z}$, such as considered in Fig.~\ref{fig:Nperp0}.
To this end, we first recall that for probe photons propagating in constant-crossed and plane-wave backgrounds the criterion $\tilde\beta=j\pi$ is met by the polarization eigenmodes, and thus no $\perp$-polarized signals are induced; cf., e.g., \cite{Bialynicka-Birula:1970nlh,Brezin:1971nd}.
Also note that in these cases momentum conservation requires that the wave vector of the probe is not changed for the quasi-elastic signal component, which is therefore emitted exactly in forward direction at $\vartheta=0$.  
At zeroth order in the paraxial approximation, the laser fields in \Eqref{eq:El,Bl} are characterized by the same vector structure as plane waves: at each space-time point the electric and magnetic fields have the same strength and are perpendicular to each other as well as to the beam axis.
An important difference is the space-time dependent envelope localizing the laser fields in all coordinates, which translates into finite energy and momentum bandwidths.
These in turn facilitate quasi-elastic signal components emitted at nonzero values of $\vartheta$.
For slowly varying fields with $\{w_0,\tau\}\gg\lambdabar_{\rm C}$ as considered throughout this work, these bandwidths scale with $\{\lambdabar_{\rm C}/w_0,\lambdabar_{\rm C}/\tau\}\ll 1$ and hence are quite small.
Because of their different propagation direction, the signal photons with $\vartheta\neq0$ and $\tilde\beta=j\pi$ are naturally no longer confined to a specific polarization eigenmode, and a $\perp$-polarized signal can arise.
The combination of these facts clarifies both the emergence and the smallness of a non-vanishing $\perp$-polarized signal for $\tilde\beta=j\pi$ at LO in the paraxial approximation: The above polarization restrictions forbid a $\perp$-signal in the kinematically favoured forward direction $\vartheta=0$. While such a signal becomes possible for $\vartheta\neq0$, it receives an effective suppression with $\vartheta\ll1$ in comparison to situations where $\tilde\beta\neq j\pi$.
At the same time,  the additional longitudinal field components in \Eqref{eq:El,Bl} at NLO in the paraxial approximation that are parametrically suppressed by $\theta_\ell\ll1$ immediately facilitate $\perp$-polarized signal components for $\vartheta\neq0$ and $\tilde\beta=j\pi$ as well.
As the would-be dominant LO component $\sim\vartheta^0$ to the $\perp$-polarized signal photon number in forward direction vanishes identically for the special polarization choice of $\tilde\beta=j\pi$, in this case there is no a priori reason why the contributions arising from \Eqref{eq:El,Bl} at linear order in the paraxial approximation $\sim\theta_\ell$ should be subleading with respect to those at zeroth order.

\begin{figure}
  \centering
  \includegraphics[width=0.9\linewidth]{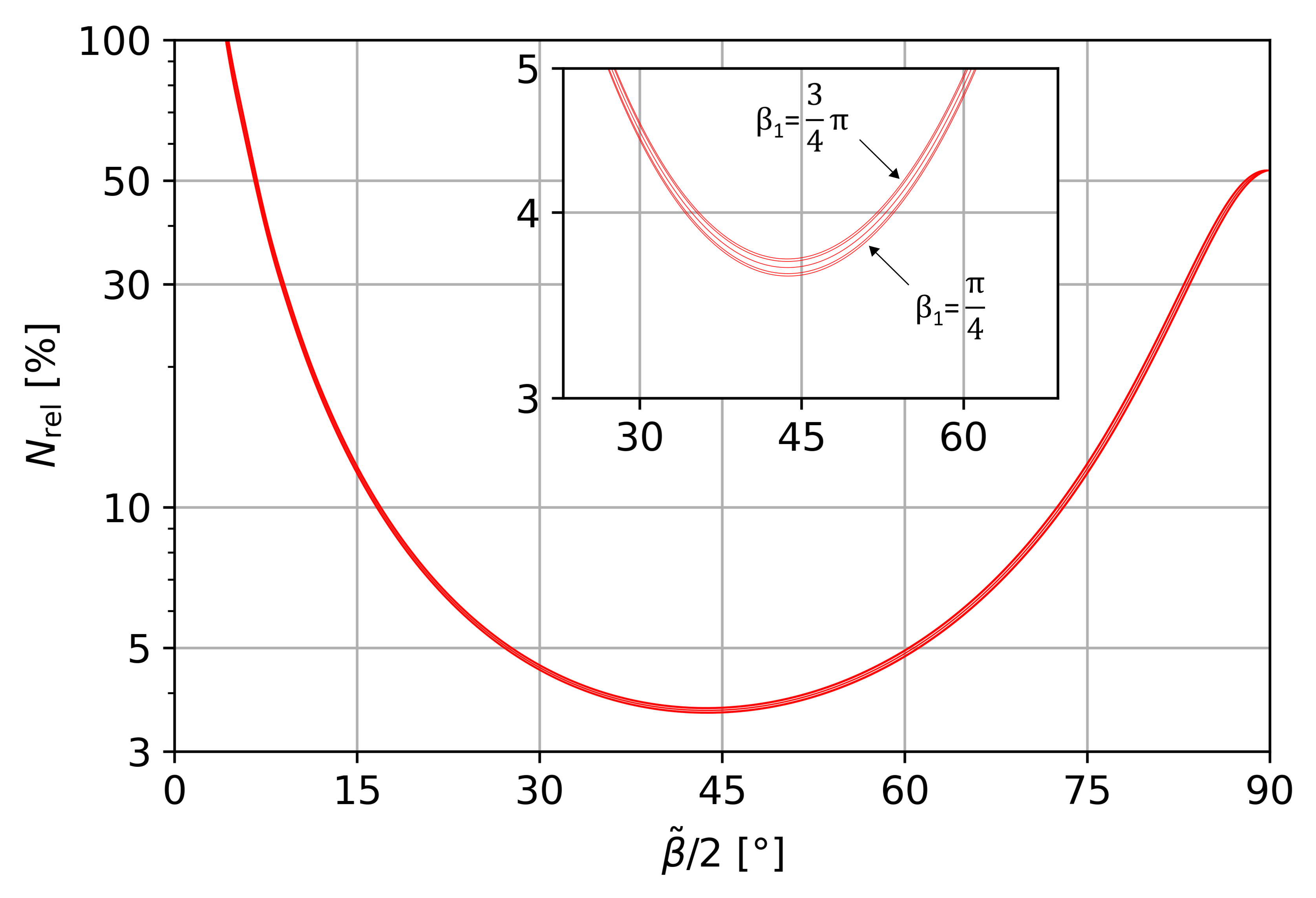}
\caption{Relative deviation $N_{\rm rel}=|N_{\perp,{\rm NLO}}-N_{\perp,{\rm LO}}|/N_{\perp,{\rm NLO}}$ of the $\perp$-polarized signals at LO and NLO in the paraxial approximation determined from \Eqref{eq:Nperp} as a function of the sum of the polarization angles $\beta_\ell$ of the colliding beams $\tilde\beta/2$ for $\tau=25\,{\rm fs}$, $w_{0,\ell}=2\lambda_\ell=1.6\,\upmu{\rm m}$ and $\vartheta_{\rm coll}=\vartheta_{\rm coll}|_{\rm max}=126^\circ$. The mild explicit dependence on $\beta_1$ is encoded in the width of the displayed line. It is essentially only visible in the inset zooming into the range $30^\circ\leq\tilde\beta/2\leq60^\circ$.}
\label{fig:Nrel_beta1}
\end{figure}

\begin{figure}
  \centering
  \includegraphics[width=0.9\linewidth]{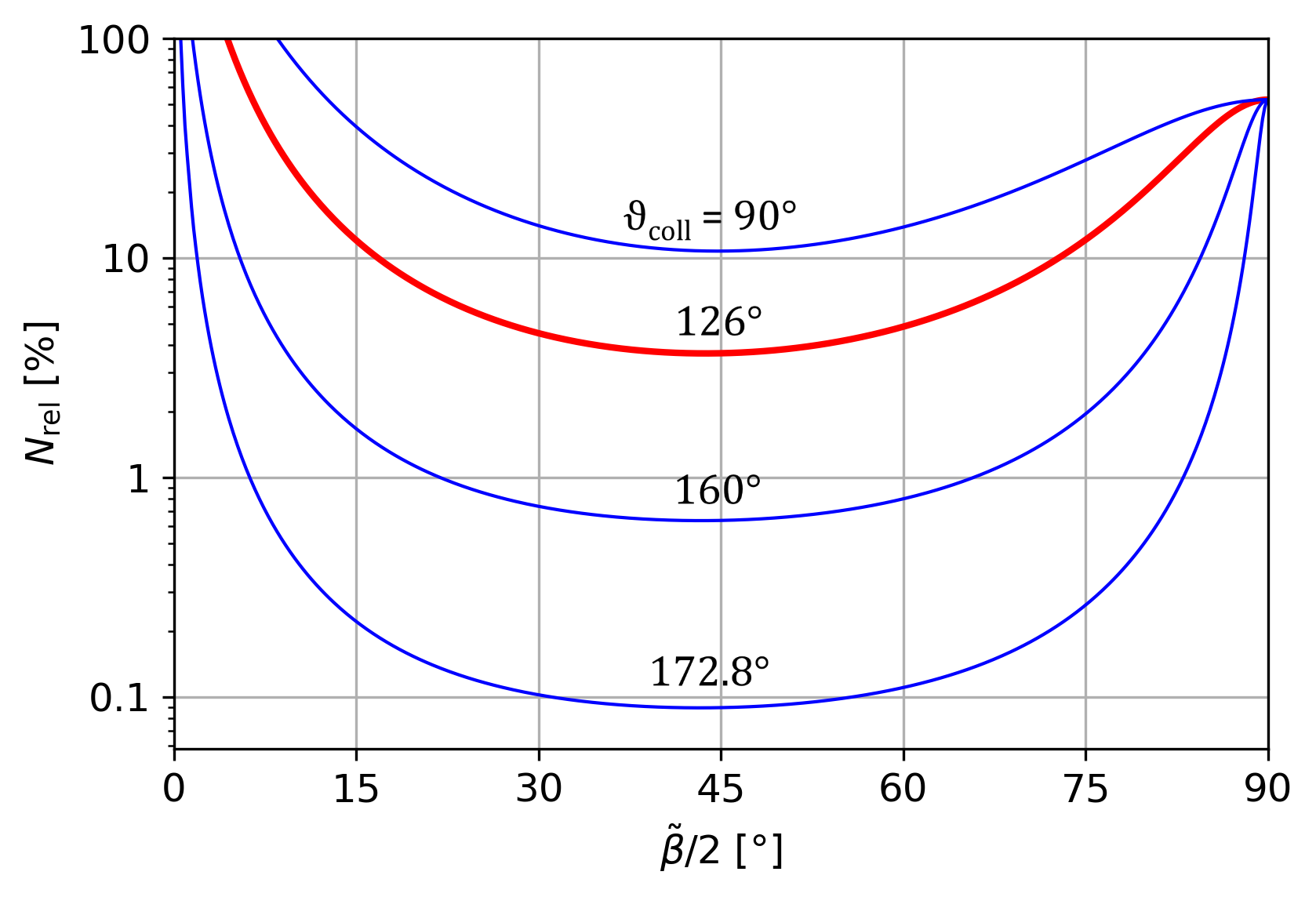}
\caption{Relative deviation $N_{\rm rel}=|N_{\perp,{\rm NLO}}-N_{\perp,{\rm LO}}|/N_{\perp,{\rm NLO}}$ of the $\perp$-polarized signals at LO and NLO in the paraxial approximation determined from \Eqref{eq:Nperp} as a function of the sum of the polarization angles $\beta_\ell$ of the colliding beams $\tilde\beta/2$ for $\tau=25\,{\rm fs}$, $w_{0,\ell}=2\lambda_\ell=1.6\,\upmu{\rm m}$ and $\beta_1=\pi/2$.
The displayed curves are for different collision angles $\vartheta_{\rm coll}\geq90^\circ$ yielding sizable signals. The thick red curve is also highlighted in Fig.~\ref{fig:Nrel_beta1}, and the lowermost one is for the largest collision angle compatible with our analytic approximation~\eqref{eq:IRLAcollangle1}.}
\label{fig:Nrel_thetacoll}
\end{figure}

Finally, we aim at briefly highlighting and quantifying the impact of higher order contributions of the paraxial approximation on $N_\perp$ beyond the exceptional cases where $\tilde\beta=j\pi$ with $j\in\mathbb{Z}$.
To this end, we study the dependence of the relative deviation $N_{\rm rel}=|N_{\perp,{\rm NLO}}-N_{\perp,{\rm LO}}|/N_{\perp,{\rm NLO}}$ of the $\perp$-polarized signals at LO and NLO in the paraxial approximation determined from \Eqref{eq:Nperp} as a function of the parameter $\tilde\beta$, or equivalently the sum of polarization angles $\beta_1+\beta_2=\tilde\beta/2$ of the colliding beams. Symmetry allows to map all cases with $\tilde\beta/2>90^\circ$ onto the domain $0^\circ\leq\tilde\beta/2\leq90^\circ$.
The case of parallel polarized laser fields with $\beta_1=\beta_2=\pi/2$ studied in Fig.~\ref{fig:Nperp0} then corresponds to $\tilde\beta/2=0^\circ$ and the one with $\beta_1=2\beta_2=\pi/2$ in Fig.~\ref{fig:Nperp45} is mapped onto $\tilde\beta/2=45^\circ$.
Aside from $\tilde\beta$, the perpendicular polarized signal~\eqref{eq:Nperp} also features an explicit dependence on $\beta_1$ via Eqs.~\eqref{eq:A4} and \eqref{eq:B2} in the appendix. For our example parameters this dependence is very mild. See Fig.~\ref{fig:Nrel_beta1}, which especially highlights the dependence on this parameter.
Figure~\ref{fig:Nrel_thetacoll} studies the relative deviation $N_{\rm rel}$ for several different choices of collision angles $\vartheta_{\rm coll}\geq90^\circ$ relevant for experiment. We note that the relative deviation tends to further increase towards smaller values of $\vartheta_{\rm coll}$. This is in line with expectations: the LO paraxial approximations in particular cannot reproduce the phenomenon of signal photon self-emission from a single beam \cite{Blinne:2018nbd} which becomes increasingly relevant towards $\vartheta_{\rm coll}\to0^\circ$. 
In all considered cases, the $N_{\rm rel}$ reaches its minimum for $\tilde\beta/2=45^\circ$.
Figures~\ref{fig:Nrel_beta1} and \ref{fig:Nrel_thetacoll} clearly demonstrate that, aside from the special cases fulfilling $\tilde\beta/2=j\pi/2$ with $j\in\mathbb{Z}$ where the NLO contribution inevitably needs to be accounted for to reliably predict the $\perp$-polarized signal, higher-order contributions to the paraxial approximation become relevant even for the case of $\tilde\beta/2=45^\circ$ if a certain precision goal is to be met.

\section{Conclusions and Outlook}\label{sec:concls}

In the present work we focused on the $\perp$-polarized quasi-elastic quantum vacuum signal induced in the collision of two linearly polarized high-intensity laser beams colliding at a generic collision angle.
Using this scenario as an illustrative example, we highlighted that while the conventional leading order paraxial beam model allows for quantitatively accurate predictions in most cases, it may also fail and significantly overestimate the yield of polarization-flipped signal photons.
This happens for the special cases where the colliding laser beams are polarized either parallel or perpendicular to each other.
Here, we clarified the origin of this deficiency and devised a strategy to overcome it. This allowed us to obtain quantitatively accurate closed-form expressions valid for all possible (linear) laser beam polarizations.
Our approach can be readily and systematically extended to higher orders in the paraxial approximation, and be generalized to the collision of more than two laser beams.
We are convinced that a well-controlled analytical approach as put forward here will be essential for understanding and unveiling the physical processes underlying the quantum vacuum signatures predicted by first-principles numerical approaches.

\acknowledgments

This work has been funded by the Deutsche Forschungsgemeinschaft (DFG) under Grant No. 416607684 within the Research Unit FOR2783/2.

\begin{widetext}

\appendix

\section{Coefficients determining the directional emission characteristics of the signal}\label{sec:nperp}

The coefficients introduced in \Eqref{eq:photondensity2ndorder} are given by
\begin{equation}\label{eq:nperpAux}
\begin{split}
    n_{\perp}^{(1)} =\, & m_{\perp}^{(1,0)} + m_{\perp}^{(1,1)}\theta_1\omega_1w_{0,1} \,, \\
    n_{\perp}^{(2)} =\, & m_{\perp}^{(2,0)} + \frac{\sin^2\vartheta_{\rm coll}}{b(1)}\Big[m_{\perp}^{(2,1)}\theta_1\omega_1w_{0,1}  + m_{\perp}^{(2,2)}(\theta_1\omega_1w_{0,1})^2 \Big]\,,
\end{split}
\end{equation}
where we defined
\begin{align}
    m_{\perp}^{(1,0)} =\, & -(1-\cos\vartheta_{\rm coll})\sin\vartheta_{\rm coll} \biggl[2\Bigl(\frac{11}{3}-\cos\tilde\beta\Bigr)\sin\varphi + \biggl(1+\frac{b\bigl((\frac{w_{0,2}}{\bar w_0})^2\bigr)}{b(3)}\biggr) \cos\varphi \sin\tilde\beta \biggr]\sin\tilde\beta\,, \\
    m_{\perp}^{(1,1)} =\, & \frac{\sin\vartheta_{\rm coll}}{b(1)}\biggl[\Bigl(\frac{w_{0,2}}{\bar w_0}\Bigr)^2\frac{b(1)}{3}\Bigl(\frac{11}{3}+\cos\tilde\beta\Bigr)\sin\varphi - b \bigl(\tfrac{1}{3}(\tfrac{w_{0,2}}{\bar w_0})^2\bigr) \cos\varphi\sin\tilde\beta\biggr]\sin\tilde\beta\,, \\
    m_{\perp}^{(2,0)} =\, & \frac{2}{9}\sin^2\vartheta_{\rm coll}\bigl(65-33\cos\tilde\beta\bigr)\sin^2\varphi \nonumber\\
    & + \biggl[\frac{1}{2}(1-\cos\vartheta_{\rm coll})^2\sin\bigl(2(\varphi-\beta_1)\bigr) + \biggl(\frac{1}{2}+\frac{b\bigl((\tfrac{w_{0,2}}{\bar w_0})^2\bigr)}{b(3)}\biggr) \sin^2\vartheta_{\rm coll}\sin(2\varphi)\biggr]\Bigl(\frac{11}{3}-\cos\tilde\beta\Bigr)\sin\tilde\beta \nonumber\\
    & +\biggl\{\biggl[3-\frac{1}{4}\biggl(11-\Bigl(\frac{w_{0,2}}{\bar w_0}\Bigr)^2 \biggr)\biggl(1-\frac{b\bigl((\frac{w_{0,2}}{\bar w_0})^2\bigr)}{b(3)}\biggr) + \frac{3}{2}\biggl(1-\frac{b\bigl((\frac{w_{0,2}}{\bar w_0})^2\bigr)}{b(3)}\biggr)^2\biggr]\sin^2\vartheta_{\rm coll}\cos^2\varphi \nonumber\\
    & \quad\quad-\biggl[1+\frac{1}{4}\biggl(2+\Bigl(\frac{w_{0,2}}{\bar w_0}\Bigr)^2\frac{b(1)}{b(0)}\biggr)\biggl(1-\frac{b\bigl((\frac{w_{0,2}}{\bar w_0})^2\bigr)}{b(3)}\biggr) \biggr] \sin^2\vartheta_{\rm coll} - \frac{b(\tfrac{3}{2})}{b(3)}\big(1-\cos\vartheta_{\rm coll}\big) \biggr\}\sin^2\tilde\beta\,, \label{eq:A4}\\
    m_{\perp}^{(2,1)} =\, & \biggl(1-\frac{b\bigl((\frac{w_{0,2}}{\bar w_0})^2\bigr)}{b(3)}\biggr)  \bigg[\frac{1}{3}\Bigl(\frac{w_{0,2}}{\bar w_0}\Bigr)^2 b(1)\sin\varphi \Bigl(\frac{11}{3}+\cos\tilde\beta\Bigr) - b\bigl(\tfrac{1}{3}(\tfrac{w_{0,2}}{\bar w_0})^2\bigr)\cos\varphi\sin\tilde\beta\bigg]\cos\varphi\sin\tilde\beta \nonumber\\
    & +\biggl[\frac{b\bigl(\frac{1}{3}(\frac{w_{0,2}}{\bar w_0})^2\bigr)}{2}-\frac{1}{2}\frac{b(0)}{1+\cos(\vartheta_{\rm coll})} + 2b\bigl(\tfrac{1}{3}(\tfrac{w_{0,2}}{\bar w_0})^2\bigr)\cos^2(\varphi)\biggr]\sin^2\tilde\beta \nonumber\\
    & +\sin(2\varphi)\bigg[\Bigl(\frac{11}{3}-\cos\tilde\beta\Bigr)b\bigl(\tfrac{1}{3}(\tfrac{w_{0,2}}{\bar w_0})^2\bigr) - \Bigl(\frac{11}{3}+\cos\tilde\beta\Bigr)\frac{1}{3}\Bigl(\frac{w_{0,2}}{\bar w_0}\Bigr)^2b(1) \bigg]\sin\tilde\beta \nonumber\\
    & -\frac{1}{6}\Bigl(\frac{w_{0,2}}{\bar w_0}\Bigr)^2b(1)\sin^2\varphi\Bigl(\frac{56}{9}+\sin^2\tilde\beta\Bigr)\,, \\
    m_{\perp}^{(2,2)} =\, & \bigg[ \frac{1}{6}\Bigl(\frac{w_{0,2}}{\bar w_0}\Bigr)^2b(1)\sin\varphi\Bigl(\frac{11}{3}+\cos\tilde\beta\Bigr) - \frac{b\bigl(\frac{1}{3}(\frac{w_{0,2}}{\bar w_0})^2\bigr)}{2}\cos\varphi\sin\tilde\beta\bigg]^2\,.
\end{align}

\section{Coefficients determining the integrated number of signal photons}\label{sec:Nperp}

In \Eqref{eq:Nperp} we introduced
\begin{align}\label{eq:NperpAux}
    N^{(2)}_{\perp} =\, & M_\perp^{(2,0)} + \Bigl(\frac{w_{0,2}}{\bar w_0}\Bigr)^2 \frac{\sin^2\vartheta_{\rm coll}}{b(1)} \Big[  M_\perp^{(2,1)}\theta_1\omega_1w_{0,1}  + M_\perp^{(2,2)} (\theta_1\omega_1w_{0,1})^2 \Big]\,,
\end{align}
with
\begin{align}
    M_\perp^{(2,0)} =\, & \frac{2}{3}\frac{b\bigl((\frac{w_{0,2}}{\bar w_0})^2\bigr)}{b(3)} \sin^2\vartheta_{\rm coll}(65-33\cos\bar{\beta}) + \frac{b(0)}{8}  \biggl(1-\frac{b\bigl((\frac{w_{0,2}}{\bar w_0})^2\bigr)}{b(3)}\biggr) \sin(2\beta_1)\Bigl(\frac{11}{3}-\cos\bar{\beta}\Bigr)\sin\bar{\beta}  \nonumber\\
    & +\biggl\{ \Bigl[2\Bigl(\frac{w_{0,2}}{\bar w_0}\Bigr)^2-3\Bigr]\sin^2\vartheta_{\rm coll} \nonumber\\ 
    & \quad\quad+ \frac{3}{4}\biggl(1-\frac{b\bigl((\frac{w_{0,2}}{\bar w_0})^2\bigr)}{b(3)}\biggr) \biggl[ 2\biggl(1-\frac{8}{3}\Bigl(\frac{w_{0,2}}{\bar w_0}\Bigr)^2\biggr) + \frac{1}{3}\Bigl(\frac{w_{0,2}}{\bar w_0}\Bigr)^2\biggl(1+\frac{1}{3}\Bigl(\frac{w_{0,2}}{\bar w_0}\Bigr)^2\biggr)\Bigl(1-\frac{b(3)}{b(0)}\Bigr)\biggr]\sin^2\vartheta_{\rm coll} \nonumber\\
    &\quad\quad + \frac{3}{4}\biggl(1-\frac{b\bigl((\frac{w_{0,2}}{\bar w_0})^2\bigr)}{b(3)}\biggr)^2 \sin^2\vartheta_{\rm coll} \biggl[2+\Bigl(\frac{w_{0,2}}{\bar w_0}\Bigr)^2\Bigl(2+\frac{b(1)}{b(0)}\Bigr)\biggr]  \nonumber\\
    & \quad\quad-\frac{b(\tfrac{3}{2})}{b(3)}(1-\cos\vartheta_{\rm coll}) \biggl[\Bigl(\frac{w_{0,2}}{\bar w_0}\Bigr)^2+ 3\frac{b\bigl((\frac{w_{0,2}}{\bar w_0})^2\bigr)}{b(3)} \biggr]\biggr\}\sin^2\tilde\beta\,,  \label{eq:B2}\\
    M_\perp^{(2,1)} =\, & \biggl\{\biggl(1+\frac{b\bigl((\frac{w_{0,2}}{\bar w_0})^2\bigr)}{b(3)}\biggr)b\bigl(\tfrac{1}{3}(\tfrac{w_{0,2}}{\bar w_0})^2\bigr) -\frac{130}{9}\frac{b(1)b\bigl((\frac{w_{0,2}}{\bar w_0})^2\bigr)}{b(3)} \nonumber\\
    & \quad+\frac{3}{2}\Bigl(\frac{\bar w_0}{w_{0,2}}\Bigr)^2\biggl(b\bigl(\tfrac{1}{3}(\tfrac{w_{0,2}}{\bar w_0})^2\bigr)-\frac{b(0)}{1+\cos\vartheta_{\rm coll}}\biggr) \biggl(\frac{1}{3}\Bigl(\frac{w_{0,2}}{\bar w_0}\Bigr)^2+\frac{b\bigl((\frac{w_{0,2}}{\bar w_0})^2\bigr)}{b(3)}\biggr) \biggr\}\sin^2\tilde\beta\,, \\
    M_\perp^{(2,2)} = \, & \frac{1}{4b(1)} \biggl[\frac{1}{3}\Bigl(\frac{w_{0,2}}{\bar w_0}\Bigr)^2 \frac{b^2(1)b\bigl((\frac{w_{0,2}}{\bar w_0})^2\bigr)}{b(3)}\Bigl(\frac{11}{3}+\cos\tilde\beta\Bigr)^2 + b^2\bigl(\tfrac{1}{3}(\tfrac{w_{0,2}}{\bar w_0})^2\bigr)\sin^2\tilde\beta\biggr]\,.
\end{align}

\vspace{-4mm}
\end{widetext}

\end{document}